
%
\headline={\ifnum\pageno=1\firstheadline\else
\ifodd\pageno\rightheadline \else\leftheadline\fi\fi}
\def\firstheadline{MIT-CTP-2258 \hfill hep-ph/9311373}
\def\rightheadline{\hfil}
\def\leftheadline{\hfil}
	\footline={\ifnum\pageno=1\firstfootline\else\otherfootline\fi}
\def\firstfootline{\rm\hss\folio\hss}
\def\otherfootline{\hfil}

\font\twelvesl=cmti10 scaled\magstep 1
\font\twelvebf=cmbx10 scaled\magstep 1
\font\twelverm=cmr10 scaled\magstep 1
\font\twelveit=cmti10 scaled\magstep 1

\font\tenbf=cmbx10
\font\tenrm=cmr10
\font\tenit=cmti10

\font\ninerm=cmr9

\parindent=1.5pc
\hsize=6.0truein
\vsize=8.5truein
\nopagenumbers

%
%
\def\phihat{ \hat{\phi} }
\def\fone{ {}^1\!\!{f} }
\def\ftwo{ {}^2\!\!{f} }
\def\phione{ {}^1\!\!{\phi} }
\def\phitwo{ {}^2\!\!{\phi} }
\def\rhonaught{ \rho_{0} }
\def\meffsq{ m_{eff}^2 }
\def\moneeffsq{ m_{1 eff}^2 }
\centerline{\tenbf PARAMETERISATION (AND GAUGE) INVARIANCE}
\baselineskip=16pt
\centerline{\tenbf OF THE}
\baselineskip=16pt
\centerline{\tenbf TRANSITION TEMPERATURE\footnote{$^*$}
{{\ninerm This paper reports upon work which was undertaken in
collaboration with R. Kobes and G. Kunstatter.}}}
\baselineskip=16pt
\vglue 0.8cm
\centerline{\tenrm P.F. KELLY}
\baselineskip=13pt
\centerline{\tenit Center for Theoretical Physics,}
\baselineskip=12pt
\centerline{\tenit Massachusetts Institute of Technology,}
\baselineskip=12pt
\centerline{\tenit CAMBRIDGE  MA  02139, U.S.A.}
\vglue 0.8cm
\centerline{\tenrm ABSTRACT}
\vglue 0.3cm
{\rightskip=3pc
 \leftskip=3pc
 \tenrm\baselineskip=12pt\noindent
The so-called Unitary Gauge Puzzle is re-examined
in the light of a set of gauge dependence identities discovered by
Kobes, Kunstatter and Rebhan.
The ``puzzle'' is discovered to arise as an artifact of the
gauge-variant and off-shell nature of the Effective Potential.
An explicit analysis of the scalar sector of the Abelian Higgs model
shows that the correct physical result is
obtained in all cases provided that the calculation is performed
self-consistently, and the result is evaluated on-shell.
\vglue 0.6cm}

\vfil
\twelverm
\baselineskip=14pt

\leftline{\twelvebf 1. Introduction}
\vglue 0.3cm

The fact that symmetries which are spontaneously broken at
zero temperature may be restored at a sufficiently large finite
temperature is by now well-appreciated$^{1,2,3}$.
The so-called Unitary Gauge Puzzle (UGP) in the context of the Abelian
Higgs model arose almost immediately upon the initial realisation that
symmetry restoration was possible$^{2,3}$.

A methodology was proposed for the estimation of the temperature at
which symmetry restoration occurs.
The {\twelvesl transition} or {\twelvesl critical} temperature, $T_{c}$, is
characterised by the vanishing of a suitable order-parameter
(it is implicitly assumed that the phase transition is of second-order).
The effective mass of the Higgs field fulfills this role.
In the standard method proposed for this analysis, the
effective thermal mass is determined perturbatively through
construction of the temperature dependent Effective Potential,
$$
\meffsq = m^2 +
2 { {\partial^2 V^{\beta} ( \phihat ) }\over{ \partial \phihat {}^2 }}
\Biggm\vert_{\phihat = 0}  \quad .
\eqno{(1)}
$$
The Effective Potential is the generator of 1PI (vertex) functions
evaluated at zero external four-momentum.
It is a function of $\phihat$, the translationally invariant
(constant) field.
The $\beta = 1/T$ superscript indicates
that attention is being paid to the temperature-dependent contributions;
the vacuum ({\twelveit viz.}, $T = 0$)
parts comprise the definition of $m^2$.
The second derivative generates the thermal contribution to the 1PI
two-point function, evaluated at zero external four-momentum, and has
a natural interpretation as a correction to the mass-squared.
The ultimate evaluation of this quantity is about the zero-field
configuration corresponding to the symmetric (restored) minimum:
$\phihat = 0$, although the calculation takes place in the broken
phase.

In the Abelian Higgs model, where
the objects in the theory are a complex scalar field, $\Phi$, and a
$U(1)$ local gauge field, $A_{\mu}$,
the Euclidean Lagrangian density is:
$$
{{\cal L}}_{E} = ( {{\cal D}}_{\mu} \Phi )^{\dag} ( {{\cal D}}^{\mu} \Phi )
+ \mu^2 \vert \Phi \vert^2
- {{\lambda}\over{3!}} ( \vert \Phi \vert^2 )^2
- {1\over4} F_{\mu\nu} F^{\mu\nu}
+ {{\cal L}}_{{ {\hbox{{\rm Gauge}}}\atop{\hbox{{\rm Fixing}}} }}  \quad .
\eqno{(2)}
$$
Note that the coefficient of the quadratic term has the
appropriate (``opposite'') sign for spontaneous symmetry breaking (SSB).
Implementing the programme sketched above is a straightforward task, but
there remains the detail of choosing a parameterisation of the scalar
field.

One natural choice is to decompose $\Phi$ into its real
and imaginary parts,
$$
\Phi = {1\over{\sqrt{2}}} [ \phi^{R} + i \phi^{I} ] \quad .
\eqno{(3)}
$$
As this parameterisation maintains the manifest (power-counting)
renormalisability of the model, the set of variables $\{ \phi^{R},
\phi^{I}, A_{\mu} \}$ has been dubbed
``renormalisable gauge.''\footnote{$^{*}$}
{{\ninerm We regret this usage because
it confuses true gauge issues with those of parameterisation.}}
Symmetry breaking is assumed to occur such that the
$\phi^{R}$ field is shifted by a vacuum expectation value (VEV),
while the quantum fluctuations
of $\phi^{I}$, and $A_{\mu}$ are about zero.
One-loop calculation of the order-parameter and insistence upon its
vanishing at $\beta_{c}$ leads to the following relation, used to
estimate $\beta_{c}$:
$$
0 = \mu^2 - {1\over{12 \beta_{c}^2}} \bigg[ 3 e^2 + {{\lambda}\over{2}}
+ {{\lambda}\over{6}} \bigg] + O( \beta_{c}^{-1} ) \quad .
\eqno(4)
$$

Yet another natural choice for the parameterisation of $\Phi$ is an
expression in terms of modulus, $\rho$, and angle
${{\theta}/{\rhonaught}}$,
{\twelveit viz.},
$$
\Phi = {1\over{\sqrt{2}}} \ \rho \ e^{i {{\theta}/{\rhonaught}} }
\quad .
\eqno{(5)}
$$
The angular argument is scaled by a dimensionful factor so that the
fields $\rho$ and $\theta$ have the same canonical dimension as $\Phi$.
The most natural choice of scale is $\rhonaught$, the zero temperature
VEV of the modulus field.
It so happens that the ``hidden symmetry'' of the system enables the
Higgs mechanism to operate, and the $\theta$-field may be eliminated
by a redefinition of the vector-gauge field $A_\mu$.
As this parameterisation makes manifest the true degrees of freedom
of the model, it has been dubbed ``unitary gauge''.
The quantum fluctuations of the modulus field are shifted away from
zero by $\rhonaught$, while those of the combined vector-gauge-Goldstone
mode are about zero.
One-loop calculation of the order-parameter, and the demand that it
vanishes at the critical temperature leads to an estimate of $\beta_{c}$
from:
$$
0 = \mu^2 - {1\over{12 \beta_{c}^2}}
\bigg[ 3 e^2 + {{\lambda}\over{2}} \bigg]
+ O( \beta_{c}^{-1} ) \quad .
\eqno{(6)}
$$

Surprisingly, equations $(4)$ and $(6)$ do not agree.
This is the UGP.
Expression $(4)$ is deemed to be correct.

There have been several proposed resolutions to this
perplexing problem which may be assigned to two schools of thought.
The first advocates some means by which the unitary calculation may
be ``fixed'' so as to obtain the correct result.
Ueda$^4$ proposed to modify the method by adding and
subtracting self-energy terms and evaluating these on the mass-shell
as determined at tree-level.
Ueda, however, was not able to provide a compelling rationale for his
refinement of the method (other than that he was able to get agreement
between the two parameterisations at one-loop order).
Arnold, Braaten, and Vokos (ABV)$^5$ suggested that the problem arose as
a consequence of a breakdown of the loop-expansion and were able to
obtain consistent results by an order-by-order reformulation in powers
of temperature.
Their analysis indicated that higher-order terms of the form
$(T/{T_{c}})^{2n}$  were inevitable, and although they were able to
verify explicitly that all of the $T^4$ terms arising from two-loop
diagrams cancelled precisely, they had no reason to believe that
the contributions at higher powers of $T$ would similarly conspire to
vanish.

Thus, ABV ended their analysis in the second school among those who
despair of ever calculating physical quantities accurately in the
unitary gauge$^6$.
The common reasoning for this opinion is as follows:  unitary gauge is
not manifestly renormalisable.
The naive power-counting arguments fail, and hence
higher-order terms in the series can dominate
over lower-order ones.
Thus, there is an intrinsic inaccuracy when
calculating in unitary gauge.
However, in spite of appearances,
the Abelian Higgs model {\twelvesl is} renormalisable.

It might seem then, that we are in the unpalatable
situation in which the physical properties that we extract from a
model depend crucially upon how we choose to parameterise
the model.
Fortunately, this is not the case and a more compelling resolution
to the UGP is presented in the rest of this paper.

\vglue 0.6cm
\leftline{\twelvebf 2. The Revised Programme for Estimation of the
Critical Temperature}
\vglue 0.3cm

Motivated by issues of gauge invariance, Kobes, Kunstatter, and Rebhan
derived a set of gauge dependence identities within the context of the
Effective Action formalism$^7$.
It was quickly realised that these formal identities were
sufficiently general to encompass parameterisation dependence as well
as gauge dependence (the two are interrelated).
Their result may be paraphrased as follows:
{\twelvesl
In a self-consistent series approximation to a physical quantity,
as expressed through the Effective Action/1PI formalism,
the results are gauge and parameterisation independent at each order
in the series when evaluated ``on-shell''}.
The issue of self-consistency and the exact nature of ``on-shell''
are discussed below.
There are two additional caveats which must be mentioned.
First, these algebraic identities may break down when
divergent quantities are encountered.
Second, the identities are not a panacea to theorists:  there is no
guarantee of the accuracy of a calculated result, only of its
gauge and parameterisation invariance.

With these considerations in mind, we advocate a revised programme for
the estimation of $T_{c}$:
\item{1).}{Choose as order-parameter the physical $\meffsq$ defined by
the pole of the propagator ({\twelveit i.e.} ``bare plus self-energy'').}
\item{2).}{Determine the self-energy corrections self-consistently
within a suitable perturbative scheme.}
\item{3).}{Calculate and self-consistently put on-shell the
temperature-dependent contributions.
Estimate $T_{c}$ from the vanishing of the order-parameter.}
\par

In the present case, the loop expansion constitutes a valid series
approximation.
Although it is conventional to regard the loop-expansion as an expansion
in powers of $\hbar$, we choose to introduce another
dimensionless loop-counting parameter $l$.
The formality of the series, and the order-by-order
gauge and parameterisation invariance are thus made more manifest.

The subject of self-consistency shall now be addressed.
The equation for the pole position is:
$$
0 = S^{-1} = P^2 + m^2 + \Pi ( P^2)  \quad .
\eqno{(7)}
$$
The self-energy (vacuum polarisation) is expanded in loops,
$$
\Pi ( P^2 ) = \Pi_{(0)} ( P^2 ) + l \, \Pi_{(1)} ( P^2 )
+ l^2 \, \Pi_{(2)} ( P^2 ) + \quad \ldots  \quad ,
\eqno{(8)}
$$
where the subscripts $(n)$ denote the loop-order of the specified
self-energy contribution, and there is no tree-level term
($\Pi_{(0)} ( P^2 ) \equiv 0 $).
The value of $P^2$ which solves the pole equation, $(7)$,
may be expanded in a series,
$$
P^2 = P_{(0)}^2 + l \, P_{(1)}^2 + l^2 \, P_{(2)} + \quad \ldots \quad ,
\eqno{(9)}
$$
leading to a consistent iterated set of equations:
$$
\eqalignno{
P_{(0)}^2 &= -m^2
&(10a)
\cr
P_{(1)}^2 &= - \Pi_{(1)} ( P_{(0)}^2 ) = - \Pi_{(1)} ( -m^2 )
&(10b)
\cr
P_{(2)}^2 &= \ldots = - \Pi_{(2)} ( -m^2 ) +
{ {\partial \Pi_{(1)} (x)}\over{\partial x} } \Biggm\vert_{- m^2}
\Pi_{(1)} ( -m^2 )
&(10c)
\cr
\ldots &{\hbox{{\twelveit etc. }}} \ldots \quad .
\cr
}
$$
Note that all quantities are evaluated on the tree-level mass-shell,
so as to be properly ordered in the perturbative expansion.

\vglue 0.6cm
\leftline{\twelvebf 3. Application to the UGP}
\vglue 0.3cm

Let us first observe, along with many other researchers, that the
essential aspects of the UGP lie entirely in the scalar sector.
In the interests of clarity, we shall restrict our attention to the
complex scalar $\Phi^4$ model with a global (ungauged) $U(1)$
symmetry.
The transition temperature will be estimated by application of the
programme discussed above.
The one-loop temperature-dependent parts will be determined in the
context of the Imaginary Time Formalism (ITF)$^{8,9}$.
The Lagrangian under consideration is:
$$
{{\cal L}}_{E} = ( \partial_{\mu} \Phi )^{\dag} ( \partial^{\mu} \Phi )
+ \mu^2 \vert \Phi \vert^2
- {{\lambda}\over{3!}} ( \vert \Phi \vert^2 )^2  \quad .
\eqno{(11)}
$$
Note that this is not a gauge theory and that the quadratic coefficient
is appropriate for SSB.

The issue of parameterisation choice arises once again.
Let us adopt a {\twelvesl one-parameter family}
of parameterisations of the complex scalar field:
$$
\eqalign{
\Phi = {1\over{\sqrt{2}}} [ \phi^{R} + i \phi^{I} ] \quad , \quad
\Phi^{R} &= (1-\epsilon) \, \phione +
\epsilon \, \phione \cos ( \phitwo / \rhonaught )  \quad ,
\cr
\Phi^{I} &= (1-\epsilon) \, \phitwo +
\epsilon \, \phione \sin ( \phitwo / \rhonaught )  \quad ,
\cr
}
\eqno{(12)}
$$
where, as before, the scale introduced for convenience is taken to be
$\rhonaught$, the zero temperature VEV.
Scrutiny of $(12)$ reveals that for $\epsilon = 0$, the cartesian
(renormalisable) parameterisation is obtained;
for $\epsilon = 1$, the polar (unitary) parameterisation results;
while for $0 < \epsilon < 1$, some
form of interpolating parameterisation ensues.

The Lagrangian $(11)$ may be recast in terms of
$\{ \phione , \phitwo \}$, with the $\epsilon$ dependence subsumed into
the coefficients.
SSB is affected by giving the entire VEV to $\phione$,
$$
\phione = \rhonaught + \fone  \quad , \quad \phitwo = 0 + \ftwo \quad ,
\eqno{(13)}
$$
where $\{ \fone , \ftwo \}$ are the quantum (fluctuating) fields.
Writing the Lagrangian in terms of $\{ \fone , \ftwo \}$ amounts to
shifting the fields in the conventional manner.

Propagators and vertices (Feynman rules) may be read from
${{\cal L}}_{E}$.
As we are seeking to determine the one-loop self-energy,
we need only keep vertices up to fourth-order in the fields.
In general, for $\epsilon \ne 0$, vertices of arbitrarily high-order
arise through the expansion of the $\sin$ and $\cos$ terms in $(12)$.
The details of this construction are thoroughly discussed
elsewhere$^{10}$.
In addition, it proves vital to include contributions from the
functional measure (a jacobean term arises from the transformation of
field variables from
$\{ \Phi , \Phi^{*} \} \rightarrow \{ \fone , \ftwo \}$), so as to
properly dispose of spurious $T^4$ terms which arise at one-loop
as a consequence of derivative interactions.
The quartic divergences which one might expect to find in the
vacuum (zero temperature) sector of the model are precisely
cancelled as well$^2$.
Renormalisation involves only the vacuum contributions to the
self-energy and is affected through the addition of
appropriate counterterms.

The upshot of this analysis is a rather complicated general expression
for the thermal contributions to the self-energy of the Higgs field:
$\fone$.
The Goldstone boson mode, $\ftwo$, can be studied also, and
consistent results are obtained.
The expression for the order-parameter,
$$
\moneeffsq = m_{1}^{2}
- \Pi^{(11)}_{ {{\hbox{TOTAL}}\atop{\hbox{thermal}} }} (P^2)
\biggm\vert_{\hbox{on-shell}}  \quad ,
\eqno{(14)}
$$
can be expanded in the high temperature limit ($\beta \rightarrow 0$).
Of course, this may only be done {\twelvesl after} analytic continuation
back to real values of the external momentum in the ITF.
In addition, there are the familiar criteria enforcing tree-level stability
which relate the coupling, masses and $\rhonaught$ so as to produce a
zero tree-level tadpole term (equivalently, yielding a massless
Goldstone boson).
These may be applied, leading to the almost final (not yet put
``on-shell'') result:
$$
\moneeffsq = 2 \mu^2 - { {\lambda}\over{9 \beta^2} }
\bigg\{ { 1 + {1\over8} \big(1 + P^2 / m_{1}^{2} \big)
   \Bigl[ ( 2 \epsilon - 1/2 )^2 - {1\over4} \Bigr] } \bigg\}
+ O( \beta^{-1} )   \quad .
\eqno{(15)}
$$
At first glance, this formula looks disastrous.
It was our intention to show parameterisation
{\twelvesl independence}, and yet our
result appears to vary with $\epsilon$.
Fortunately, this is not the case.
Setting the result ``on-shell'' fixes $P^2 = - m_{1}^{2}$,
causing the $\epsilon$-dependent part to appear with a vanishing
coefficient, and hence the correct equation for the critical
temperature,
$$
0 = 2 \mu^2 - { {\lambda}\over{9 \beta_{c}^2} } + O ( \beta_{c}^{-1} )
\quad ,
\eqno{(16)}
$$
is obtained for all values of $\epsilon$, {\twelveit viz.,}
for all parameterisations.
Thus the UGP has been avoided through our insistence upon using a
physical quantity for the order-parameter, and evaluating it
self-consistently on-shell.

Equation $(15)$ has more content than has yet been exposed.
For instance, when $\epsilon$ is set to zero ({\twelveit i.e.,} the
cartesian parameterisation), the term in the square brackets
vanishes identically and the correct result $(16)$ is obtained
whether the on-shell condition is enforced or not.
This is not an accident.
It stems rather from the fact that the self-energy to one-loop in
the cartesian parameterisation is independent of the external momentum
as a consequence of the absence of derivative interactions.
Thus, one is able to better understand the success of the Effective
Potential method in capturing the correct physics in the case of the
cartesian parameterisation.
Astoundingly, inspection of $(15)$ reveals another instance of this
fortuitous ability to obtain the correct result both on- and off-shell,
in the case of $\epsilon = 1/2$, the ``midway'' parameterisation.
This particular feature is most unlikely to be of any genuine
physical consequence.
Finally, through $(15)$, we are able to arrive at the ``incorrect''
result, $(6)$, as well.
We accomplish this by setting $\epsilon = 1$, selecting the
polar or unitary parameterisation, and, contrary to our
prescription, {\twelvesl setting the external four-momentum to zero}
in the evaluation of the order-parameter.
This shows unequivocally the origin and resolution of the UGP.

\vglue 0.6cm
\leftline{\twelvebf 4. Conclusion}
\vglue 0.3cm

Application of a prescription based upon the self-consistent calculation
of a physical quantity, and its subsequent on-shell evaluation, has
demonstrated that the estimate of the critical temperature, so obtained,
may be expected to be independent of the parameterisation and gauge used.
Of course, the accuracy of such a calculation must be checked by some
other means
(best undertaken in a renormalisable gauge or parameterisation).
Our results have solved the long-standing UGP by revealing it to be an
artifact of the difficulties inherent in the extraction of physical
quantities from a gauge-variant, off-shell object, such as the Effective
Potential.

\goodbreak
\vglue 0.6cm
\leftline{\twelvebf 5. Acknowledgements}
\vglue 0.4cm

I thank my collaborators:  R.L. Kobes and G. Kunstatter for the
opportunity to first present our work at this meeting.
Many thanks are due to the conference organisers for their parts in
helping to create a most enjoyable and stimulating environment.
In addition, I wish to thank R. Jackiw, M.E. Carrington, and S. Vokos
for their helpful criticisms.
This work was supported by the Natural Sciences and Engineering Research
Council of Canada, and by the Winnipeg Institute for Theoretical
Physics.

\vglue 0.6cm
\leftline{\twelvebf 6. References}
\vglue 0.4cm

\medskip
\itemitem{1.} D.A. Kirzhnits and A.D. Linde,
{\twelveit Phys. Lett.} {\twelvebf B42} (1972) 471.
\itemitem{2.} L. Dolan and R. Jackiw,
{\twelveit Phys.  Rev.} {\twelvebf D9} (1974) 3320.
\itemitem{3.} S. Weinberg,
{\twelveit Phys.  Rev.} {\twelvebf D9} (1974) 3357.
\itemitem{4.} Y. Ueda,
{\twelveit Phys.  Rev.} {\twelvebf D23} (1981) 1383.
\itemitem{5.} P. Arnold, E. Braaten and S. Vokos,
{\twelveit Phys.  Rev.} {\twelvebf D46} (1992) 3576.
\itemitem{6.} see for example,
M. Chaichain, E.J. Ferrer and V. de la Incera,
{\twelveit Nucl. Phys.} {\twelvebf B362} (1991) 616.
\itemitem{7.} R. Kobes, G. Kunstatter and A. Rebhan,
{\twelveit Nucl. Phys.} {\twelvebf B355} (1991) 1.
\itemitem{8.} J.I. Kapusta,
{\twelveit Finite Temperature Field Theory}
(Cambridge University Press, Cambridge, England, 1989).
\itemitem{9.} N.P. Landsmann and Ch. G. van Weert,
{\twelveit Phys.  Rep.} {\twelvebf 145} (1987) 141.
\itemitem{10.} P.F. Kelly, R.L. Kobes and G. Kunstatter,
in preparation.
\bye